\documentstyle[12pt,prl,aps,epsf,preprint,tighten]{revtex}

\begin{document}

\draft
\preprint{HKBU-CNS-9815}
\title{Non-Monotone Characteristic of Spectral Statistics in
the Transition between Poisson and Gauss}

\author{ 
Hiroshi Hasegawa$^1$, Baowen Li$^2$, Jian -Zhong Ma$^3$, and Bambi 
Hu$^{2,4}$}
\address{ 
$^1$ Research Center of Quantum Communications\\
Tamagawa University, Machida, Tokyo 194, Japan\\
$^2$Department of Physics and Centre for Nonlinear Studies\\ Hong Kong 
Baptist University, Hong Kong, China\\
$^3$Max-Planck Institut f\"{u}r Kern Physik,
 D-69117 Heidelberg, Germany\\
$^4$ Department of Physics, University of Houston, TX77204, USA}
\date{\today}
\maketitle

\begin{abstract}
We have computed the spectral number variances of an extended random matrix ensemble
predicted by Guhr's supersymmetry formula, showing a non-monotone increase of the
curves that arises from an \,"overshoot\," of the two-level correlation function
above unity.  On the basis of the most general form of $N$-level joint distribution
that meets sound probabilistic conditions on matrix spaces, the above characteristic
may be attributed to the {\it attractiveness} of the pair potential in long range($E
>$ Thouless energy) of the underlying level gas. The approach of level dynamics
indicates that the result is \,"anti-screening\," of the level repulsion in
short-range statistics of the usual random matrix prediction until the joint level
distribution undergoes a phase transition(the Anderson transition).
\end{abstract}
  
\pacs{PACS numbers: 05.45.+b, 05.20.-y, 05.40.+j, 72.15.Rn, 02.50.-r}

There have been considerable efforts in solid-state and random matrix
theories(RMT) to formalize metal-insulater transition phenomena as regards
the pertinent electron energy level statistics, which seek a powerful and
unified method to generalize the standard Gaussian ensembles initiated by
Wigner, Dyson and Mehta\cite{Mehta} (see a comprehensive review on the 
recent development\cite{GMW97}).

Guhr has studied a method based on supersymmetry\cite{Guhr89}, and 
recently obtained
a formula for computing the two-level correlation function $X_2(r)$ for a
Hamiltonian system $ H = H_0 + \alpha H_1$ with a parameter
$\lambda$($\lambda=\lambda(\alpha)$) to describe a transition from
regularity to chaos\cite{Guhr96a,Guhr96b}; $0 \le \lambda \le \infty$.  
It is assumed that
$\lambda = 0$(also $\alpha = 0$) represents the Poisson regularity for the
system $H_0$ alone (which is subject to the Poisson statistics), and
$\lambda = \infty$ the Gaussian unitary ensemble (GUE) for the system $H_1$
alone (which is subject to the GUE statistics).  Namely, in Refs. 
\cite{Guhr96a,Guhr96b}, a
double integration representation of the $X_2(r,\lambda)$ function is
presented: the one involving a single Bessel function 
$J_1(z)$\cite{Guhr96b} is shown to converge for $\lambda \to \infty$ to the
well-known GUS correlation function
$1- \left( \frac{{\rm sin}\pi r}{\pi r} \right)^2$, and
for the lowest non-zero term of the small $\lambda$ 
\begin{equation} 
X_2(r, \lambda) = \frac{r}{\lambda} \int_0^\infty {\rm exp}
\left( -k^2/2 \right) {\rm sin}(rk/2)dk \quad ({\rm valid \, \, 
for} \, \, \lambda \simeq 0.1 \, {\rm or \, \, smaller} ).
\end{equation}

Our concern in this report is not any inquiry about the supersymmetry
basis of the formula nor its derivation, but a consequence of what Guhr
called "overshoot" of the quantity $X_2(r,\lambda)$ above unity, i.e.

\begin{equation}{\rm a \, \, range \, \, of \,} \, \, r \, \, {\rm exists, 
\, \, where\,\, } X_2(r,\lambda) > 1.
\label{X2b} 
\end{equation}
\noindent
This can best be visualized by plotting $X_2(r,\lambda)$ against $r$: 
here in Fig.1, we calculate four curves
$\lambda =$ 0.1, 5, 10 and $\infty$, and compare them
with the corresponding curves which are provided by another formula of
Gaudin's model due to us\cite{HaMa98} (Some feature of this model will be 
discussed later). 

Since the unfolded scale is used always in the present formulas and
figures (which also ammounts to a change of the perturbation parameter
$\alpha$ to $\lambda$), inequality (\ref{X2b}) implies the negativeness 
of the
corresponding cluster function $Y_2(r,\lambda)$\cite{Mehta} so that 
$$
Y_2(r,\lambda) \equiv 1-X_2(r,\lambda) < 0 \, \, {\rm in \, \, the \, \,
same \, \, range}.  \eqno{(2')} $$ 
This causes a significant modification
of the structure of the level statistics for such systems from the
standard RMT.  We would like to point out the following two remarks. 

1. {\bf Number variance curve} based on the well-known formula 
\begin{equation}
\Sigma^2(s,\lambda) = s - 2\int_0^s(s-r)Y_2(r,\lambda)dr.  \quad (
Y_2(-r,\lambda) = Y_2(r,\lambda). )
\label{S2}
\end{equation}
\noindent
It can be seen readily from Eq.(\ref{S2}) that 
$\frac{d^2}{ds^2}\Sigma^2(s) =
-Y_2(s) <0$, which means that a $\Sigma^2(s)$ curve has an inflection
point at the zero of $Y_2$: more precisly, the increasing behavior of the
curve changes its second derivative from minus to plus at the inflection
point.  

2. {\bf N-level joint distribution} which we assume to be of the form 
\begin{equation}
P(x_1,x_2,..,x_N) = \frac{1}{Z_\beta}{{\rm exp}\left[-
\beta\sum_{j<k}\phi(x_j-x_k)\right]}, \qquad \beta = 1, \, \, 2\,\, {\rm 
and} \, \, 4. 
\label{PX}
\end{equation}
Here, the sum $\sum_{j<k} \phi(x_j-x_k)$ in the right hand side of eq.(4) to
represent the interaction between the levels is common in most of extended RMT
nowadays, and we shall state its axiomatic basis in {\bf 1st aspect} of Balian's
strategy below.  \\ \hspace*{3mm} The cluster expansion theory for imperfect
classical gases in statistical mechanics tells us that, if the binary potential for
a pair of the molecules is positive (i.e. everywhere repulsive), the second-order
cluster function must be non-negative at least for the low density limit where it is
approximated by the minus of the Mayer function, $1-e^{-\beta\phi(r)}(\beta > 0)$
(\cite{Toda92} and a further account therein).  Guhr's example of the smallest
parameter value can be regarded as such a limiting situation so that the existence
of the overshoot of Guhr's correlation function on the $r$-axis has a significance
that, when the joint distribution of his model is expressed in the form of
Eq.(\ref{PX}), the potential $\phi$ for a pair of levels $x_j$ and $x_k$ must have
an attractive portion somewhere in the configuration space ($x_j, x_k$) of the
levels. {\it Does this mean a level attraction rather than the level repulsion of
the standard RMT sense?} \\
\hspace*{3mm}      
The possibility of long-range level attraction has been discussed by Jalabert,
Pichard and Beenakker\cite{JPB93} for actual disordered metals.  We shall discuss
this question, referring to another recent paper by Weinmann and Pichard \cite{WP96}
who observed the behavior of non-monotone increase of the number-variance curves in
a selected actual matrix ensemble and analyzed the data on their advocation of
"Gaussian matrix ensemble with preferential basis"(cf.\cite{PS94}).  Before going,
we present a numerical and graphical confirmation of the non-monotone behaviors of
the theoretical $\Sigma^2$ curves predicted by Guhr's formula (1) and that from a
related approximation (see Eq. (4.3) in Ref.\cite{FGM98}).

As can be seen in Fig.2(a), the non-monotone characteristic is common for
all the parameter values, although the overshoot becomes obscure in
Fig.1(a) quickly as $\lambda$ increases.  So, it is important to fix the
range of $\lambda$ on which the non-monotone character remains to exist,
in particular, to ask its existence for large
$\lambda$'s.  The latter question 
has been answered affirmatively by Fram, Guhr, and 
M\"{u}ller-Groeling\cite{FGM98}, who have provided a numerically tractable 
formula to replace Eq.(1) for $X_2(r,\lambda)$.  Moreover, they 
have shown a further simplification of the formula which turns out to be   
a divergent perturbation(i.e. starting from $\lambda = \infty$) by which
$\Sigma^2(s)$ for $\lambda \gg 1 $ can be understood easily. 

Leaving the actual graphs of confirmation to the paper\cite{FGM98}, we argue the
existence of overshoot for $\lambda \gg 1$ by means of their reduced formula: $$
X_2(r,\lambda) = X_2^{GUE}(r)+X_2^{(1)}(r,\lambda)  +X_2^{(2)}(r,\lambda) +{\rm
(correction)}, $$ where the term (correction)  vanishes faster than $r^{-2}$ for $r
\to \infty$. In terms of the cluster function $Y_2(r,\lambda)$ with neglect of such
corrections,

\begin{equation}
Y_2(r,\lambda) = \left( \frac{{\rm sin}\pi r}{\pi r} \right)^2 -
X_2^{(1)}(r,\lambda) - X_2^{(2)}(r,\lambda), 
\label{Y2}
\end{equation}
where 
$$ X_2^{(1)}(r,\lambda) =
\frac{1}{\pi}\frac{\pi\lambda^2}{(\pi\lambda^2)^2 + r^2}, \quad
\int_{-\infty}^{\infty}X_2^{(1)}(r,\lambda)dr = 1, \eqno{(5a)} $$ and $$
X_2^{(2)}(r,\lambda) = \frac{1}{2\pi^2}\frac{r^2 - (\pi\lambda^2)^2 }
{((\pi\lambda^2)^2 + r^2)^2}, \quad
\int_{-\infty}^{\infty}X_2^{(2)}(r,\lambda)dr = 0.  
\eqno{(5b)} $$
\noindent
It shows that the first-order correction $X_2^{(1)}$(Breit-Wigner term) compensates
the familiar GUE part, together with the second-order correction $X_2^{(2)}$
contributing negatively to the cluster function for almost all $r$ values($|r| >
r_0$ with some small $r_0$), which 
confirms the negativeness of $Y_2(r, \lambda)$ above the
$r_0$.  At the same time,
$\int_{-\infty}^{+\infty} Y_2(r,\lambda)dr = 0$ \, for almost all
$\lambda$ values, indicating that the $\Sigma^2$ curve approaches an asymptotic
straight line that is the pure Poisson with no deviation of the coefficient from
unity.  The feature quite differs from the counter example of Gaudin's
model\cite{HaMa98}, as exhibited in Fig.2(b). 
 
To be a significance in the above two works, the authors of 
\cite{WP96} and \cite{FGM98} have
presented actual matrix ensembles which display, as computer experimental
results, the non-monotone number-variances:  this stimulates us to construct
an adequate statistical model capable of a unified description of both
types; the monotone type and the non-monotone type.  So, let us examine
the concept of Gaussian matrix ensemble with preferential basis of
Refs.\cite{WP96,PS94}.  

Our conclusion is precedently stated: the formulation as it stands is not capable of
describing the non-monotone characteristic, because their potential function $\phi$
is everywhere positive, as exemplified by Eq.(\ref{PG}) of Ref.\cite{PS94} that is
identical to \cite{HaMa98} for Fig.2(b).  However, this formulation can be revised
without loss of sound mathematical basis to include the non-monotone characteristic,
which we will discuss in the rest of the present report.

There are two aspects of the {\it Gaussian matrix ensemble with preferential
basis} to be reexamined seriously, in order to allow it with the
possibility of attractiveness of the pair potential:  the first is a
proper introduction of the {\it preferential basis}, and the second, a
right way of setting up constraints in {\it maximum entropy principle} for
the Gaussian probability density function.  Both aspects are combined to
give the most efficient realization of Balian's strategy\cite{Balian}. 
 
{\bf 1st aspect.} An element of Balian's strategy in RMT is an introduction of
Riemannian metrics on matix spaces in the form Tr$dMdM^*$ by which an information
quantity is written (his {\it postulate A}).  This is particularly relevant in the
present problem, because a fixing of the metric tensor is the starting point of the
whole subject of a Gaussian probability that is an exponential of the metric form. 
Namely,
\begin{equation} P(A) = C{\rm exp}[-\frac{1}{2}{\rm K}(A,A)], \quad {\rm K}(A,A) =
\sum_{\alpha,\beta}K_{\alpha,\beta}A_\alpha^*A_\beta \ge 0 \quad ({\rm equality \,
\, only} A=0), \end{equation} 
\noindent
where $A_\alpha$ is a tangent vector component of the metric form as the Gaussian
random variable.  We then ask what is the most general metric form defined on matrix
spaces (See a detailed discussion\cite{Hasegawa}). The answer is given by a {\it
covariant bilinear form } of any two $N \times N$ hermitian(or, unitary) matrices
$A$ and $B$ depending on another hermitian matrix $H$ to satisfy {\it representation
invariance} all together.  It is expressed as
\begin{equation} {\rm K}_H (B,A) = {\rm Re Tr}B^*{\rm C}_H (A), 
\label{Kh}
\end{equation} 
in terms of a linear superoperator ${\rm C}_H$ on $A$, and satiesfies the
covariance condition 
$$ {\rm K}_{U^*HU} (U^*BU,U^*AU) = {\rm K}_H(B,A) \quad 
{\rm for \, \, any \, \, unitary} \, \, U \in G_\beta,  
\eqno{(7a)} $$ 
where $G_\beta$ is the symmetry group associated with $\beta$.  An adequate choice
of the representation is naturally the $H$-diagonal representation ( $H_D = {\rm
diag}(x_1,..,x_N)$ by a choice of $U$ to diagonalize $H$) that yields {\it
preferential basis}:  by this choice we can rewrite (7) as
\begin{equation}
{\rm K}_H (A,A) = \sum_j c(x_j)|A_{jj}|^2 + 2\sum_{j<k} 
f(x_j,x_k)|A_{jk}|^2
\qquad {\rm with} \, \, {\rm real \, \, positive} \, \, c(x), f(x,y)
\label{Kha} 
\end{equation}
$$ {\rm and} \qquad f(x,y) = f(y,x).  \eqno{(8a)} $$

By definition of the trace operation in eq.(7), there is no actual priority of the
representation basis in our "preferential basis", as it were supposed by the saying
"$H$-diagonal representation" that might sound a kind of symmetry break[10] to
prohibit some $U \in G_\beta$. Instead, the unitary covariance (7a) allows $U$ to
cover the full group $G_\beta$ so that the Gaussian variables $\{ A_{jk} \}$ are
decorrelated from $\{ x_j \}$, and upon being integrated over these variables, the
resulting reduced probability density function depends on the eigenvalue indices
only through $\{ x_j \}$ that are mutually fully equivallent to each other(so-called
{\it identically distributed}).  \, Another characteristic of the Gaussian
probability function (6), when the form (7) is inserted (with $\alpha = jk$), is the
{\it statistical independence} of $A_\alpha$'s; the fact that the metric tensor in
the form(8) is diagonal with respect to $\alpha$.  Let us further impose the third
condition of {\it translational invariance}($x_j \to x_j + x$ makes the metric
tensor unchanged).  Then, the form (8) is further simplified such that a single
function $f(x-y)$ is enough to characterize the $N$-level joint probability function
$P(x_1,..,x_N)$ as in the form (4). (This is seen by the integration of the function
(6) over the {\it cotangent} variables defined by $\tilde{A}_{jk}=
2f(x_j-x_k)A_{jk}$.)  We then get

\begin{equation}
\phi(r) = -\frac{1}{2}{\rm log}f(r) = \frac{1}{2} {\rm log}\frac{1}{f(r)}. 
\label{phi}
\end{equation}
This is because the result of the integrations is expressed
as the square-root of the determinant of the metric tensor 
that yields
$P(x_1,..,x_N)$.  At the same time, a probabilistic meaning of the
function $f(x_j-x_k)$ is assingned to be the variance of the Gaussian
cotangent variable $\tilde{A}_{jk}$(or, its each component, if it is
complex, and quaternion depending on the multiplicity $\beta = 2, \, {\rm
and} \, 4$, respectively).    To summarize, we can say \\
{\it the most general expression for $N$-joint level distributin
$P_N$ as an integral reduction of the Gaussian form (6) 
satisfying the identicalness, statistical independence and
translational invariance must be the form (4)}.  
(An inclusion of one-level potentials is another matter.)  
\\
{\bf 2nd aspect.} It is important to remark that, in order to apply the maximum
entropy principle to assign a form to the probability function $P(\tilde{A})$
(Balian's {\it postulate B}), a Gaussian probability has a special property of its
entropy.  Namely, the Gaussian probability function $P_G(\tilde{A}_{jk})$ with the
variance $f(x_j-x_k)$ is characterized by the maximum entropy of all the probability
functions with the same variance, thus denoting the entropy functional of $P,
\langle -{\rm log}P \rangle_P$, by $H[P]$, we can express this fact as

\begin{equation}
\max_P H[P] = H[P_G] \qquad {\rm under \, \,
constraint} \, \,
{\rm Var}(\tilde{A}_{jk}) = f(x_j-x_k).  
\label{HP}
\end{equation}
(For the legitimacy of a maximum entropy principle with
non-constant constraint, see [14].)  It tells us
that the variance (\ref{HP}) must be the right quantity of constraint 
for the
present maximization problem.  Therefore, without any further information
about the matrix ensemble, we have no criterion about the degree of
attractiveness of the pair potential $\phi$ that is only related to the
form of variance $f$ as Eq.(\ref{phi}): all what is needed is the positivity $f > 0$.  

However, a powerful information can be provided by {\it level
dynamics}\cite{HaMa98}(cf.[14]), where it is shown that, if the diagonalization 
process of a
hermitian matrix $H$ (or, its unitarization $U_0 e^{iH}$) is put into a
Hamiltonian dynamics, a possible equilibrium state of this dynamical
system can be selected by the above maximization as a one-parameter family
of canonical distributions (in the sense of statistical mechanics of
Hamiltonian systems) and represented by the form 
\begin{equation} 
P_G(\tilde{A})
\propto \prod_{j<k}{\rm 
exp}\left[-\frac{1}{2f(x_j-x_k)}|\tilde{A}_{jk}|^2\right],
\quad {\rm with} \, \, f(r) = | \frac{\mu r^2}{1 + \mu r^2} | \, \, ({\rm
hermitian \, \, case}), 
\label{PG} 
\end{equation}
and hence 
$$ \phi(r) =
\frac{1}{2}{\rm log}|1 + \frac{1}{\mu r^2}|.\eqno{(11a)} $$ 
A positive
parameter value $\mu$ corresponds to Gaudin's linear-gas 
model\cite{Gaudin}, also
identical to the formula of Refs.\cite{WP96,PS94} for which $\phi$ is 
fully repulsive. To our emphasis, however, there is no {\it a priori} 
reason to restrict us to the positivity of $\mu$ (as far as $f(r)$ is retained as
positive), and let us look at the possibility of choosing a {\it negative}
$\mu$(This amounts to a sign change of Yukawa's parameter $\gamma$\cite{Yukawa}): it
makes the potential $\phi$ partially attractive such that

\begin{eqnarray}
\phi(r) =
\frac{1}{2}{\log}(\frac{1}{|\mu|r^2}-1) \quad |r| < 1/\sqrt{|\mu|}; \qquad
\frac{1}{2}{\log}(1-\frac{1}{|\mu|r^2}) \quad |r| > 
1/\sqrt{|\mu|},\nonumber\\
{\rm with \, \, the \, \, attractive \, \, range} \qquad 1/\sqrt{2|\mu|} <
|r| < \infty,
\label{phir}
\end{eqnarray} 
and at the same time, for small values of the level distance including the repulsive
range($|r| < 1/\sqrt{2|\mu|}$),
\begin{equation} 
f(r) = r^2|\mu|/(1- |\mu|r^2) \qquad |r| < 1/\sqrt{|\mu|}. 
\label{cr}
\end{equation}
(A schematic behavior of the entire potential function $\phi(r)$ can be seen in Fig.3.)
This can be interpreted as an {\it anti-screening} of the level repulsion
in short-range rather than the screening assigned to it in Gaudin's model
by the original proposal\cite{PS94}, and as the actual level attraction in the range
specified by (12) for long-range level statistics.

There exist two routes of the Wigner-Dyson Gaussian standard of electron energy
level statistics to Poisson regularity; one $\mu (= 0)$ tends to infinity on the
positive axis, and the other $\mu (= 0)$ to infinity on the negative axis, and it is
remarkable that the latter route undergoes a singularity of the level distribution
that can well be assigned to be the Anderson transition with Thouless energy $E_c =
\Delta/\sqrt{|\mu|}$($\Delta$ is the average spacing in energy scale ):  the
resulting Poisson statistics must be on localized electrons that, at the same time,
give rise to a real symmetry-breaking of the representation (the metric tensor then
depends on the indices $jk$ as $e^{-\alpha|j-k|}$). \, Existence of both repulsive
and attractive inter-level pair-wise interaction was first indicated by \cite{JPB93}
for disordered metals, and from the present viewpoint these provide a distinction
between absence and presence of the transition, and it can be tested by the
monotonicity or non-monotonicity in the number-variance curve as in Fig.2. 
\vspace{2mm}

We would like to thank T. Guhr for his communication of the latest work in
Ref. \cite{FGM98} and for a fruitful discussion on the present 
subject. BL and BH 
were supported in part by grants from the Hong Kong Research Grants 
Council (RGC) and the Hong Kong Baptist University Faculty Research Grants 
(FRG).

\newpage


\begin{figure}
\epsfxsize=14cm
\epsfbox{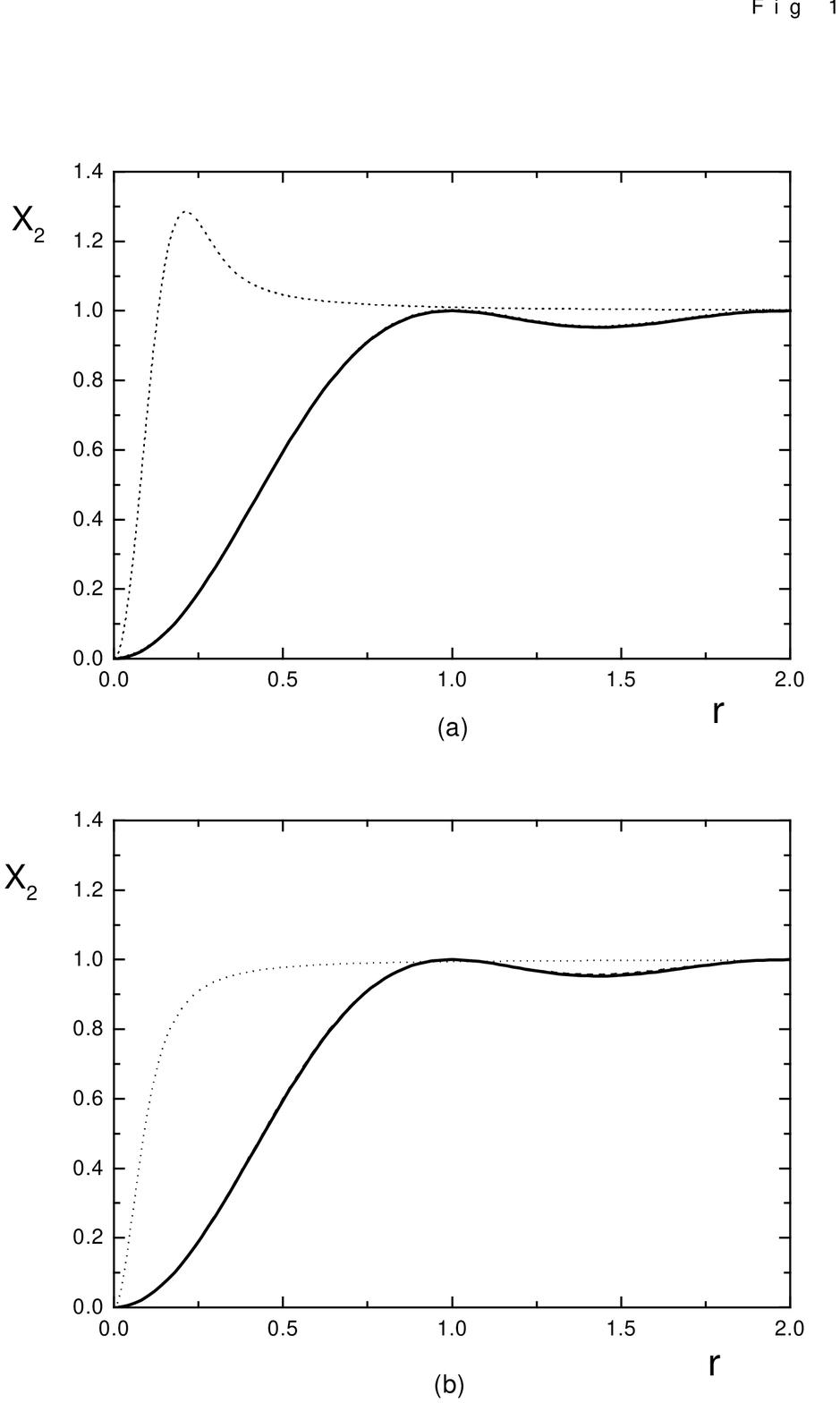}
\vspace{-1cm}
\caption
{
$X_2 (r,\lambda)$ for different 
values of the transition parameter $\lambda$ from two different theories. 
(a) Calculated from Guhr's formular for $\lambda\ll 1$  
and $\pi \lambda^2 \gg 1$. (b) Calculated from Gaudin's 
model (Eqs. (4.5) of Ref. [6] ) From the left to 
right the 
curves corresponde to $\lambda=0.1, 5, 10$ and $\infty$. The curves of 
$\lambda=$ 5 and 10 are almost indistinguishable from the GUE curve (solid 
line).
}
\end{figure}

\begin{figure}
\epsfxsize=14cm
\epsfbox{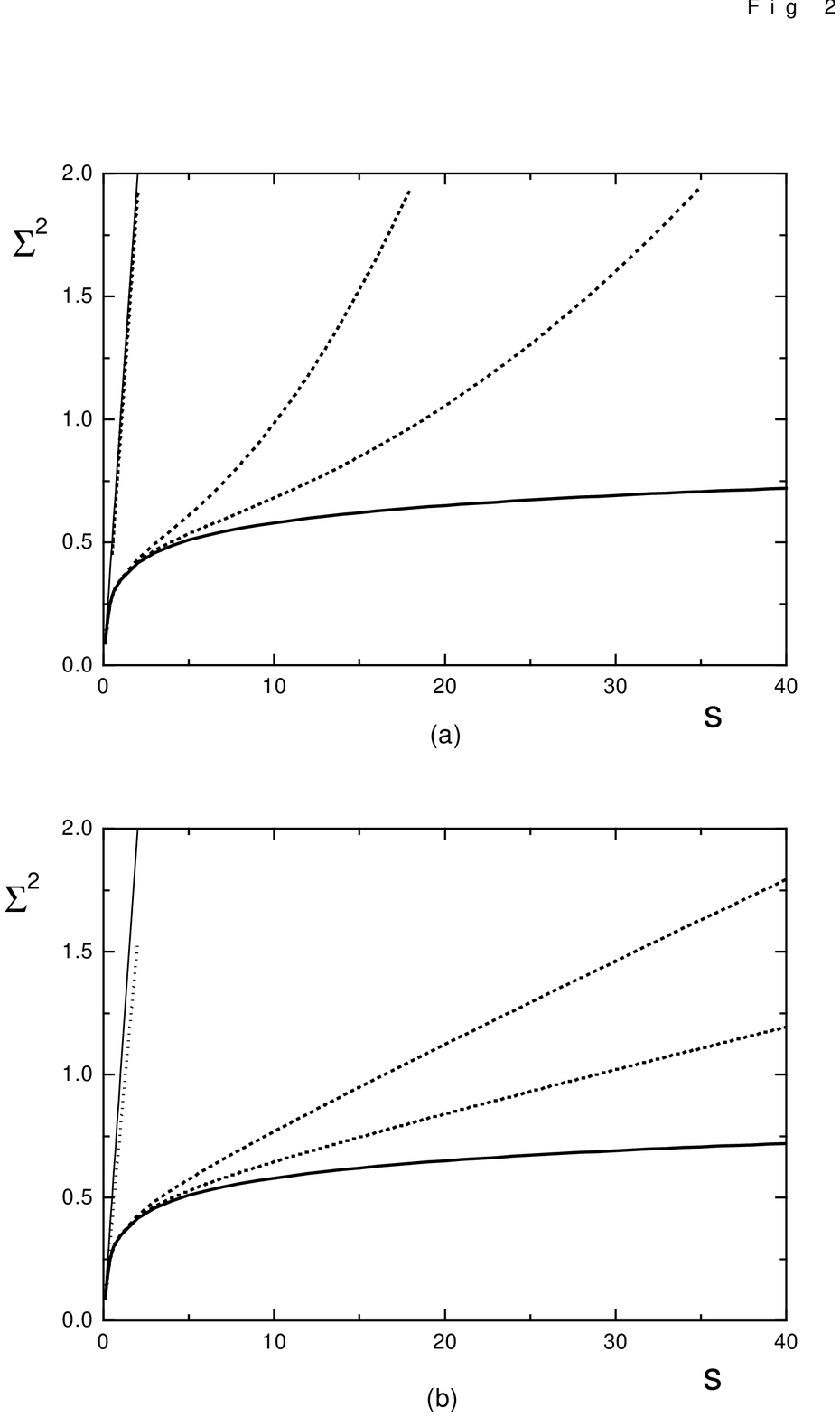}
\vspace{-1cm}
\caption
{ Number variance $\Sigma^2(s,\lambda)$ for two different models. 
(a) From Guhr's model (Eq. (4.4) of Ref. [10]). (b) From 
Gaudin's model due to Hasegawa 
and Ma [6].  The non-monotonic behavior in (a) is very obvious.
From the left to 
right the curves corresponde to the Poisson, $\lambda=0.1, 5, 10$ and 
$\infty$ (GUE), respectively. 
}
\end{figure}

\begin{figure}
\epsfxsize=14cm
\epsfbox{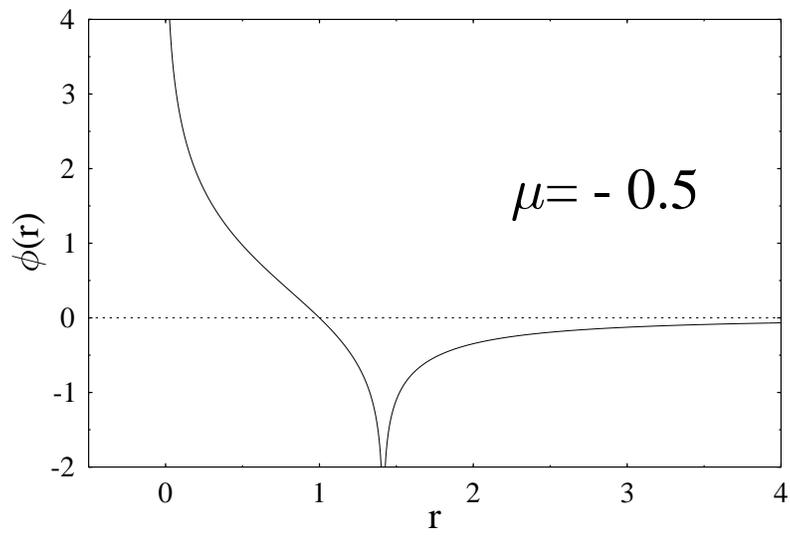}
\vspace{-1cm}
\caption
{$\phi(r)$ vs $r$ for $\mu=-0.5$ given by Eq. (11a).}
\end{figure}

\end{document}